\documentclass[reprint,nofootinbib,twocolumn,prl,superscriptaddress,showpacs,aps,prb]{revtex4-1}
\usepackage[utf8]{inputenc}
\usepackage[T1]{fontenc}
\usepackage{graphicx}
\usepackage{amsmath}
\usepackage{hyperref}

\begin{document}
\title{Microkelvin electronics on a pulse-tube cryostat\\with a gate Coulomb blockade thermometer}
\author{Mohammad Samani}
\email{m@msamani.ca}
\author{Christian P. Scheller}
\thanks{This author contributed equally to this work as the first author.}
\affiliation{Department of Physics, University of Basel, CH-4056 Basel, Switzerland}
\author{Nikolai Yurttagül}
\author{Kestutis Grigoras}
\author{David Gunnarsson}
\thanks{Present address: Bluefors Oy, Arinatie 10, 00370 Helsinki, Finland}
\affiliation{VTT Technical Research Centre of Finland Ltd, P.O. Box 1000, FI-02044 VTT Espoo,
	Finland}
\author{Omid Sharifi Sedeh}
\affiliation{Department of Physics, University of Basel, CH-4056 Basel, Switzerland}
\author{Alexander T. Jones}
\author{Jonathan R. Prance}
\author{Richard P. Haley}
\affiliation{Department of Physics, Lancaster University, Bailrigg, Lancaster LA1 4YB, UK}
\author{Mika Prunnila}
\affiliation{VTT Technical Research Centre of Finland Ltd, P.O. Box 1000, FI-02044 VTT Espoo, Finland}
\author{Dominik M. Zumbühl}
\email{dominik.zumbuhl@unibas.ch}
\affiliation{Department of Physics, University of Basel, CH-4056 Basel, Switzerland}

\begin{abstract}
Access to lower temperatures has consistently enabled scientific breakthroughs. Pushing the limits of \emph{on-chip} temperatures deep into the microkelvin regime would open the door to unprecedented quantum coherence, novel quantum states of matter, and also the discovery of unexpected phenomena. Adiabatic demagnetization is the workhorse of microkelvin cooling, requiring a dilution refrigerator precooling stage. Pulse-tube dilution refrigerators have grown enormously in popularity due to their vast experimental space and independence of helium, but their unavoidable vibrations are making microkelvin cooling very difficult. On-chip thermometry in this unexplored territory is also not a trivial task due to extreme sensitivity to noise. Here, we present a pulse-tube compatible microkelvin sample holder with on-board cooling and microwave filtering and introduce a new type of temperature sensor, the gate Coulomb blockade thermometer (gCBT), working deep into the microkelvin regime. Using on- and off-chip cooling, we demonstrate electronic temperatures as low as 224$\pm$7~µK, remaining below 300~µK for 27 hours, thus providing sufficient time for measurements. Finally, we give an outlook for cooling below 50~µK for a new generation of microkelvin transport experiments.
\end{abstract}
\maketitle

\section*{Introduction}
Several open questions in condensed matter physics are pending experimental access to lower device temperatures than what is possible today. Examples include topological ordering \cite{Hasan2010TopologicalOrdering}, electron-nuclear ferromagnets \cite{Braunecker2013TopologicalMajorana,Chekhovich2013}, quantum Hall ferromagnetic states \cite{Chesi2008QHFerromagnet}, p-wave superconductivity \cite{Mackenzie2003Sr2RuO4}, non-Abelian anyons \cite{Cooper2009}, and anomalous metallic phases in disordered superconductors \cite{Kapitulnik2019DisorderedSuperconductor}. Lower electron temperatures can also lead to longer coherence times in semiconducting and superconducting qubits \cite{Hanson2007}.

Reaching temperatures below 1~mK in nanoelectronic devices, however, poses several challenges. Commercially available dilution refrigerators offer cooling down to $\approx$5~mK at the mixing chamber. But due to weak electron-phonon coupling at such temperatures, cooling electrons in nanoelectric devices is quite difficult, requiring careful filtering and thermalization of the electrical leads \cite{Scheller2014SilverEpoxy}.

To cool to even lower temperatures, adiabatic nuclear demagnetization is the most widely used technique, and it is capable of reaching microkelvin temperatures in bulk metals \cite{Pickett1988AND,Pobell2007AND}. In this single-shot method, first the nuclear spins in an appropriate paramagnetic metal -- the nuclear refrigerator (NR) -- are polarized by applying a large magnetic field and then removing the heat of magnetization, typically by using a dilution refrigerator over a number of days. Subsequently, the field is ramped down adiabatically to a small but finite final field. Ideally, this reduces the temperature of the nuclear-spins subsystem by the same factor as the reduction in field. The heat leaking in from the environment is absorbed by the NR and slowly destroys the polarization of the nuclear spins which gives them a finite cooling power. This technique has been used very successfully and broadly, for example in producing record-low nuclear-spin temperatures down to 280~pK \cite{Knuuttila2001}. Because electrons in the NR couple directly and strongly to the nuclear spins, demagnetization can be very effective in lowering the temperature of electrons in nanoelectronic devices with 
on-chip NRs that cool the device directly.

\begin{figure*}[ht]
	\centering
	\includegraphics[width=\linewidth]{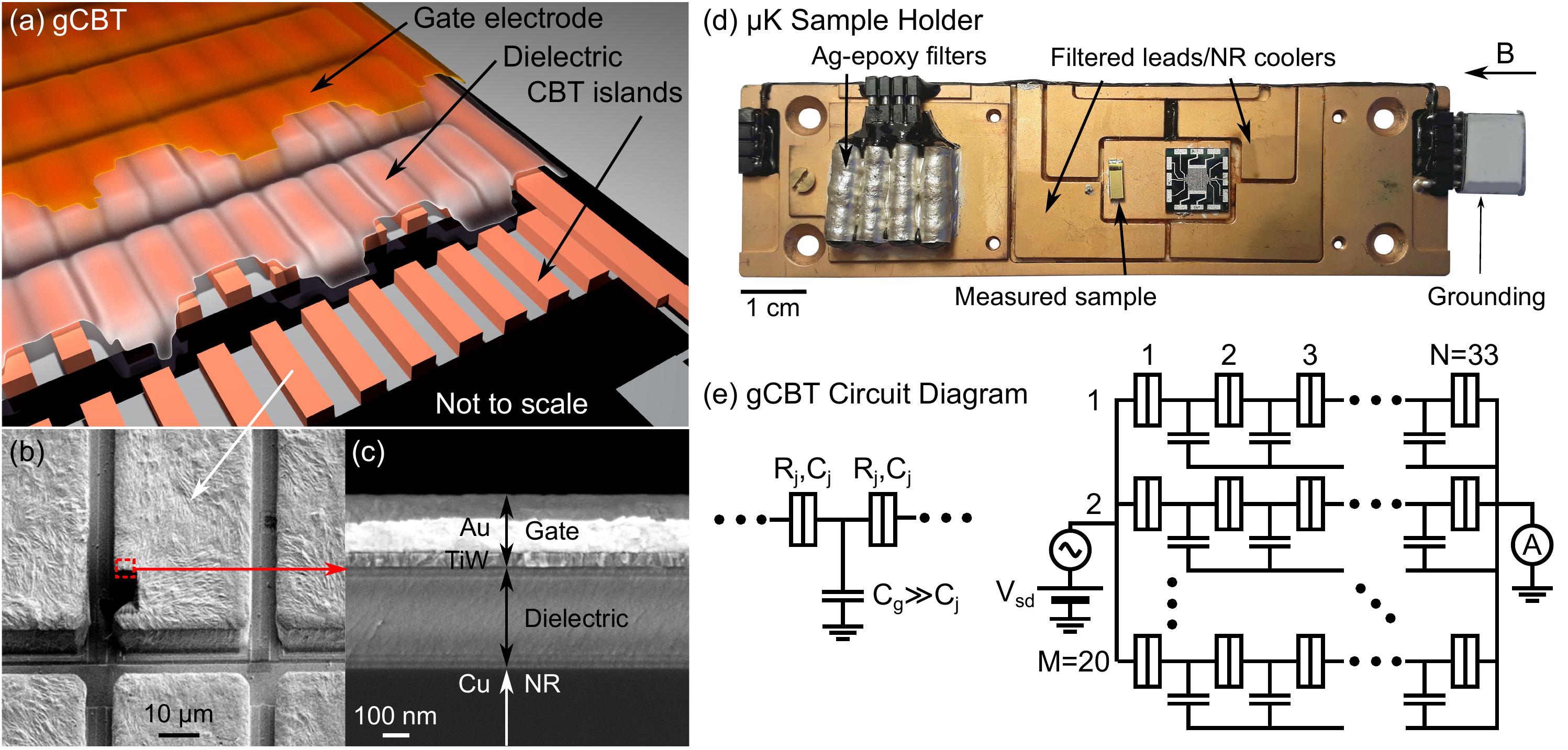}
	\caption{{\bf Enabling cooling and thermometry at microkelvin temperatures} (a) Schematic design of the gCBT. A metal gate is sputtered conformally on top of a dielectric layer covering all islands. Each island contains a block of copper, acting as NR material for cooling individual islands.
		(b) Scanning electron microscope monograph of a gCBT viewed from the top.
		(c) Coating’s cross section obtained by focused ion beam milling.
		(d) The sample box is made of gold-plated copper and is screwed to one of the macroscopic copper plates. Non-inductive silver-epoxy microwave filters \cite{Scheller2014SilverEpoxy} are placed perpendicular to the magnetic field to minimize pickup. Before the cooldown, the enclosure is completed with a copper lid screwed to the 4 small holes to block radiation, and to provide a microkelvin environment for the sample, since the NR material of the box, the bond pads, and the lid also cool due to demagnetization. The bond pads are electrically isolated from the box using cigarette paper and varnish. A grounding mechanism keeps the device protected when it is disconnected from the measurement setup.
		(e) Left: one island consists of two tunnel junctions with resistance $R_j$ and capacitance $C_j$. The island has capacitance $C_g$ to the gate. Right: the gCBT is a grid of M=20 parallel rows, with N=33 serial junctions in each row.}
	\label{fig:cbtbox}
\end{figure*}

Cryogen-free dilution refrigerators employ helium only in a closed loop and have recently become very popular. They offer a large experimental space and do not consume liquid helium -- a non-renewable resource -- hence reducing cost and eliminating disruptive helium transfers. However, the cooling power relies on a pulse tube which unfortunately causes mechanical vibrations at low frequencies that are difficult to decouple from the experiment. The vibrations, combined with the magnetic field required for the demagnetization technique, cause significant heating and thus pose an additional challenge for employing demagnetization in cryogen-free systems compared to traditional wet fridges.

Cooling electrons in nanoelectronic devices through phonons is especially difficult due to the poor electron-phonon coupling below 10~mK \cite{Bradley2016,Iftikhar2016}. Various parasitic heat leaks arising from mechanical vibrations \cite{Palma2017MagneticCooling}, microwave radiation \cite{Saira2012,Zorin1995},  heat release from materials \cite{Pobell2007AND}, and electronic noise \cite{Maradan2014} warm up electrons when they are decoupled from the mixing chamber's cooling power.

Integrating NR materials \emph{on-chip} \cite{Ciccarelli2016} places the source of cooling directly where it is most effective \cite{Bradley2017OnChip} but still leaves the chip exposed to sources of heat coming from warmer stages. This limits the lowest temperatures attainable and reduces the hold time at those temperatures. Providing a low-temperature, low-heat-leak environment for the sample is, therefore, essential in reaching ultra-low temperatures.


With this in mind, we have built a parallel network of NRs where each wire is cooled by its own separate NR in the form of a macroscopic copper plate \cite{Clark2010,Palma2017OnOff}, providing \emph{off-chip} cooling ideally suited to cool nanoelectronic samples \cite{Casparis2012,Feshchenko2015}. In addition, the chip is enclosed in a filtered copper box which itself is demagnetized (see Fig.~\ref{fig:cbtbox}(d)), thus providing a microkelvin environment for the sample. The combination of \emph{on-} and \emph{off-chip} cooling has proven quite successful, first at reaching 2.8~mK \cite{Palma2017OnOff}, and then 1.8~mK \cite{Jones2020}. These experiments were performed on a cryogen-free refrigerator and are still limited predominantly by the pulse-tube vibrations. In a wet dilution refrigerator (without a pulse tube), cooling to temperatures as low as 421~µK was recently demonstrated with indium as the NR material \cite{Sarsby2020}. 

Thermometry at low temperatures also poses several challenges, particularly in transport electronic measurements. CBTs are well established for measuring the temperature of electrons directly \cite{Pekola1994}. Due to Coulomb blockade effects, appearing when the thermal energy of electrons $k_B T$ becomes comparable or less than the charging energy $E_c$ of the quantum dot or tunnel-junction island, the differential conductance $g$ as a function of source-drain voltage exhibits a temperature-dependent minimum near zero bias, as shown in Fig~\ref{fig:calibration}(a). The width of this dip can be used as a primary thermometer for the temperature of the electrons \cite{Meschke2004} if overheating is negligible. Once $E_c$ is known, the depth $\delta g$, measured at zero bias and thus not suffering from bias heating, can be used for thermometry \cite{Casparis2012}.

In the \emph{deep} Coulomb blockade regime, $k_B T \ll E_c$, the thermal broadening of the Fermi reservoirs can be directly extracted from the conductance measurements, for example from the dependence of conductance on gate voltage or source-drain voltage. However, such measurements are very sensitive to local charges, including uncontrolled trapped charges on or near islands, or charge noise from switchers, traps, or two-level systems. These effects deteriorate the energy resolution of the measurement and introduce a potentially large temperature uncertainty. This can be avoided in the \emph{universal} Coulomb blockade regime $k_B T \gtrsim E_c$, where the conductance is independent of local charges, making the device immune to offset charges or switchers. The universal regime was shown to extend down to about $k_B T \gtrsim 0.8 E_c$ \cite{Feshchenko2013}. Thus, to measure temperatures below 1~mK, CBTs with very low charging energies of around 1~mK or lower are required. This corresponds to large on-chip capacitances, contrary to the trend in nanoscience to reduce the size and thus also the capacitance of islands. Compared to our previous work with a charging energy of 3.3~mK$\times k_B$ \cite{Palma2017OnOff}, a large decrease of charging energy, and a corresponding increase of capacitance, is needed.

In this report, we present a new type of CBT that is covered with a gate metallization layer -- termed gCBT -- depicted in Fig.~\ref{fig:cbtbox}. The metal gate increases the total capacitance of islands by a factor of 6, yet, surprisingly, decreases the charging energy by a factor of 12, down to 737$\pm$2~µK. This is because the increase in capacitance is entirely due to the improved island-to-gate capacitance, thus affecting the charge dynamics and reducing the charging energy. 
Moreover, this new type of thermometer turns out to be more immune to offset charges and nanofabrication imperfections, extending the range of validity to much lower temperatures, down to 160~µK$\approx$0.2~$E_c/k_B$, with an accuracy better than 10\%, making the gCBT an excellent microkelvin thermometer.

The device is mounted in a special sample holder which provides a microkelvin environment with very low noise and very low heat leaks as required for experiments below 1~mK, despite being mounted on a pulse-tube cooler vibrating at 1.4~Hz and higher harmonics. The holder consists of a copper box acting as a Faraday cage with a sample stage, extensive non-inductive microwave filters \cite{Scheller2014SilverEpoxy} on the measurement wires, and copper bond pads, all packaged in a mechanically rigid manner to minimize vibration effects. The copper box and bond pads together with the copper on the device are also used as NR material, providing on- and off-chip cooling. We also carefully zero the DC input voltage, first by using the second-harmonic signal from the lock-in at higher temperatures and then by short source-drain voltage scans at lower temperatures, in order to read the temperature properly and to avoid additional Joule heating. With these advances, we demonstrate electronic temperatures as low as 224$\pm$7~µK, a new record for cooling nanoelectronics. We also demonstrate that the temperature remains below 300~µK for more than 27 hours, making this technique useful for extended measurements at microkelvin temperatures.

\section*{Results And Discussions}

\subsection*{Coulomb Blockade Thermometer}
A CBT in its simplest form consists of an island enclosed between two tunnel junctions. The energy cost of adding an electron to the island is $E_c = e^2/2C_\Sigma$ where the total capacitance of the island is $C_\Sigma = 2C_j+C_g$, $C_j$ is the capacitance through a junction, and $C_g$ is the capacitance to the environment, here mostly a metal gate. When thermal energy $k_B T$ is low compared to $E_c$, the flow of electrons through the island is suppressed due to Coulomb blockade effects, appearing as a temperature-dependent dip in conductance near zero source-drain voltage bias $V_\mathrm{sd}$. 
For an array of islands and tunnel junctions in series, when $C_j \gg C_g$, these effects can be very well modelled by a master-equation approach \cite{Pekola1994,Farhangfar1997}, allowing us to convert the conductance to a temperature within some range of validity. The CBT used in this experiment is a grid consisting of N=33 tunnel junctions in series and M=20 rows connected in parallel. Using a chain of tunnel junctions in series divides the applied voltage per island by N, thus significantly reducing the effects of voltage noise. This is an important advantage considering that the voltage noise $V_\mathrm{n}$ on the islands needs to be much smaller than the temperature to be measured: $eV_\mathrm{n} \ll k_B T$. Having several chains of junctions in parallel provides a larger and therefore an easier-to-measure current.

\begin{figure}[ht]
	\centering
	\includegraphics[width=\linewidth]{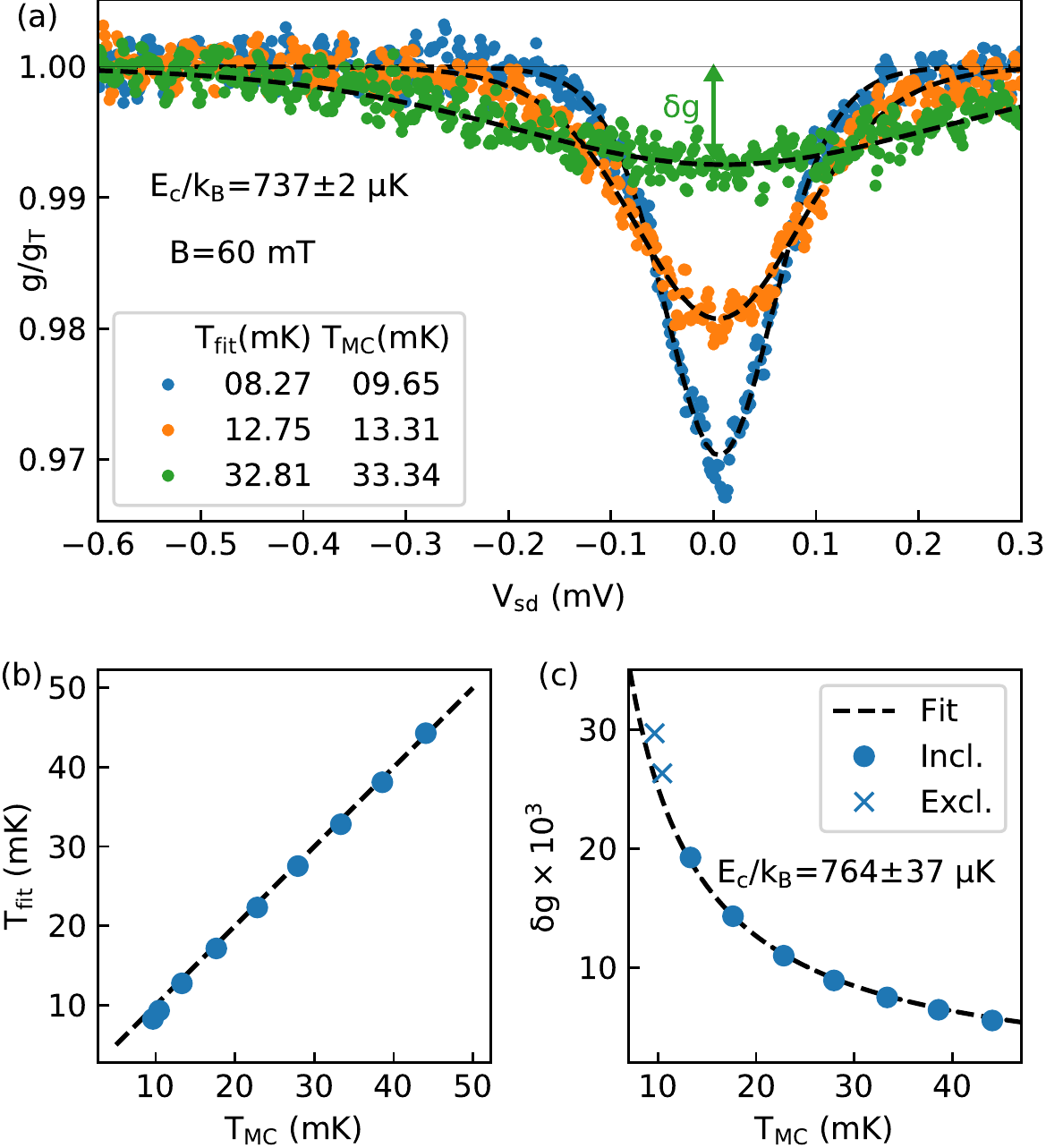}
	\caption{{\bf Characterization of the gCBT} (a) Normalized differential conductance $g/g_T$ as a function of $V_\mathrm{sd}$ at 3 representative temperatures, where $g_T$ is the high $V_\mathrm{sd}$ limit of conductance $g$. The dashed curves are extracted from a simultaneous fit of the master equation to 9 such traces, all sharing $E_c$ as a common parameter, while having individual temperature parameters, shown in the legends as $T_{\mathrm{fit}}$. (b) Comparison of temperatures from primary CBT mode $T_{\mathrm{fit}}$ and the mixing chamber RuO$_2$ thermometer reading $T_\mathrm{MC}$. Very good agreement is found.
		(c) Secondary thermometer calibration: The normalized zero-bias dip $\delta g=1-g_0/g_T$ is shown with the fit to the third order polynomial. Only the data points marked with circles are included in the fit.}
	\label{fig:calibration}
\end{figure}

In equilibrium -- in absence of bias heating -- the CBT acts as a primary thermometer, and its full-width at half-maximum is proportional to the electron temperature. A simultaneous fit of the master equation to 9 traces, 3 of which are shown in Fig.~\ref{fig:calibration}(a), at different temperatures delivers the common charging energy $E_c$=737$\pm$2~µK quite precisely. The extracted temperatures $T_\mathrm{fit}$ are compared to the mixing chamber temperature $T_\mathrm{MC}$ in Fig.~\ref{fig:calibration}(b). Very good agreement is found between the two sets. At the two lowest temperatures, the CBT reads a slightly lower value than $T_\mathrm{MC}$. We attribute this behaviour to some self-heating of the mixing chamber resistor at the lowest temperatures.

\begin{figure}[ht]
	\centering
	\includegraphics[width=\linewidth]{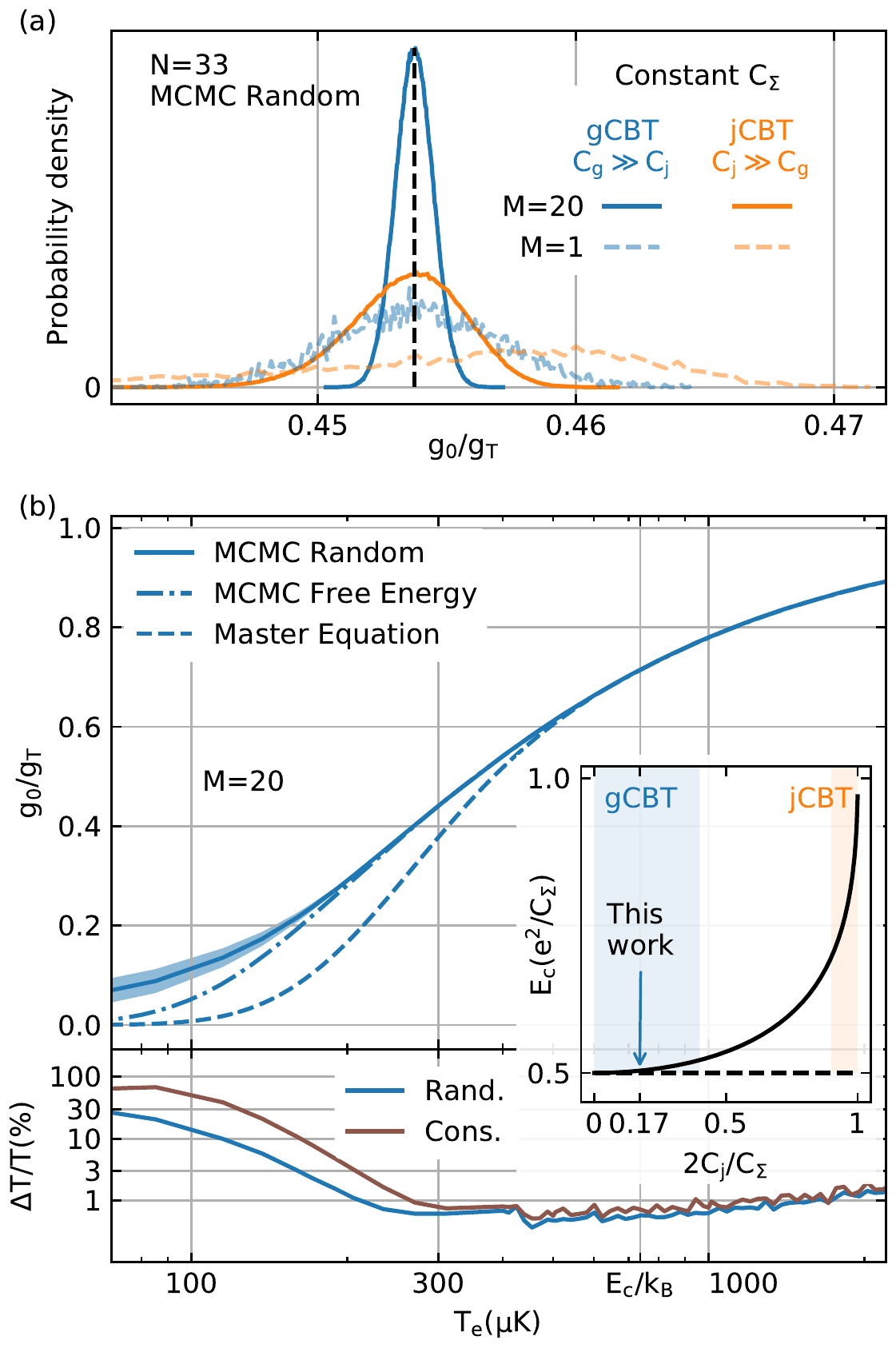}
	\caption{{\bf Simulated precision and range of the gCBT} (a) Probability distribution of normalized zero-bias conductance $g_0/g_T$ for cases as labelled with M parallel chains and N junctions in each chain. A narrower distribution corresponds to a more precise thermometer. All 4 curves have the same mean conductance (black-dashed line). The gCBT is at $T$=311~µK and the jCBT at $T$=317~µK. (b) Temperature dependence of the gCBT computed using the MCMC model, showing $g_0/g_T$ in the upper panel and the relative temperature uncertainty $\Delta T/T$ in the lower panel. 
		The shaded area around the solid line is the 3$\sigma$ uncertainty range based on conductance distributions, also shown in the lower panel (Rand.). The conservative (Cons.) relative error in the lower panel is the difference between the upper bound of the 3$\sigma$ range and the free-energy-minimized value. The inset shows the charging energy $E_c$ as a function of $C_j$ for constant $C_\Sigma$. N=2 for the dashed line and N=33 for the solid line.}
	\label{fig:mcmc}
\end{figure}

Alternatively, the CBT may be used as a secondary thermometer operating at zero bias. This avoids Joule-heating due to the applied voltage. Such heating effects can be very pronounced, particularly at the lowest temperatures, where also the zero-bias dip becomes extremely narrow, despite its broadening in the array by the number of islands $N$ in the chain. However, as a secondary thermometer, we first need to calibrate the CBT against another thermometer to obtain its charging energy. This can be done against the CBT itself at higher temperatures, shown in Fig.~\ref{fig:calibration}(a), or against the mixing chamber, in Fig.~\ref{fig:calibration}(c) where the zero-bias conductance $g_0$ is measured over a range of mixing chamber temperatures; we then fit a 3\textsuperscript{rd} order polynomial approximation \cite{Farhangfar1997} of the master equation to the data over the range $T\geq10$~mK where the mixing chamber reading is believed to be accurate, with the charging energy as the only fit parameter. The agreement with the data is very good, and the extracted charging energies are the same, within error bars, between the two modes of operation. Once $E_c$ is known, the 3\textsuperscript{rd} order polynomial is used to extract the temperature from $g_0$ for $k_B T > E_c$. For the rest of this work, we use the $E_c$ extracted from the simultaneous fitting of $V_\mathrm{sd}$ versus $g$.

The master equation is derived for a single island under the assumption that $C_g \ll C_j$, but the opposite is true for our device. Therefore we develop a simulation based on the Markov-chain Monte-Carlo (MCMC) method to capture the physics of the device, and to calculate the effect of background charge offsets on conductance. See the supplementary material for the details of the numerical method. In Fig.~\ref{fig:mcmc}(a) we compare the probability distribution of measuring a certain $g_0$ between the gCBT with $C_g \gg C_j$ (blue) and the jCBT with $C_j \gg C_g$ (orange). For one row of N=33 tunnel junctions, the probability distribution for the gCBT is a factor of 3 narrower than the jCBT. It is also more symmetric. This implies the gCBT features a range that would extend to significantly lower temperatures, or, at a fixed temperature, would provide a more precise reading, than the jCBT. Having M=20 parallel rows of tunnel junctions not only allows more current to flow through the grid, it also provides averaging over $M$ random charge offset distributions, making it even narrower compared to M=1 by more than a factor of 4 \cite{Yurttagul2021}.



In Fig.~\ref{fig:mcmc}(b), the temperature-dependence of $g_0$, normalized by $g_T$, is shown for the master equation (dashed) and compared with the MCMC simulation (solid) with the same $C_\Sigma$. Above 600~µK$\approx$0.8$E_c/k_B$, the two produce the same $g_0$, but below, the master equation consistently predicts a lower $g_0$ than the MCMC. At temperatures below 200~µK$\approx$0.25$E_c/k_B$, the uncertainty, derived from the width of the probability distribution in the MCMC model, starts to get larger, indicated by the shaded area which shows the 3$\sigma$ confidence interval around the mean $g_0/g_T$. We plot the relative error $\Delta T/T$ based on the $3\sigma$ bands on the bottom panel (solid curve), and find that the error remains surprisingly small, down to 115~µK$\approx$0.15$E_c/k_B$ with accuracy better than 10\%. This confidence interval originates from a completely random sampling of the offset charge distributions on the array of islands. Alternatively, one may limit the analysis to offset charge distributions in the MCMC simulation that minimize the free-energy functional of the CBT array. This gives the dot-dashed curve for $g_0/g_T$, which turns out to have an even smaller confidence interval from its MCMC simulations. Only below $\approx$200~µK, the zero-bias conductance produced by the two methods of MCMC start to slowly diverge, with the free-energy sampling always lying below in conductance.

The physics of the charge arrangement on such arrays is complicated because of the unknown combination of mobile charges, corresponding to the free-energy sampling, and fixed charges, corresponding to the fully random sampling. In absence of a more detailed understanding of the charging physics, we take the more conservative approach of using the fully random and free-energy-minimized sampling as the bounds of the error estimate, shown in the lower panel as the brown $\Delta T/T$ curve. Even with this rather conservative estimate, the gCBT remains better than 10\% accurate down to 160~µK~$\approx$~0.20$E_c/k_B$. This is a much improved range extending to significantly lower temperatures compared to
the jCBT, where $T\gtrsim$0.8$E_c/k_B$ \cite{Yurttagul2021}. The larger temperature range as well as the increased conductance $g_0$ above the master equation values, see Fig.~\ref{fig:mcmc}(b), are mainly due to the dominant gate capacitance which makes the electronic states of the islands essentially independent of each other. By contrast, the adjacent islands in a junction CBT capacitively couple, resulting in highly correlated islands which suppress the conductance strongly. This effectively extends the universal regime of the gate CBT to much lower temperatures, where eventually deep Coulomb blockade is setting in, suppressing the conductance to zero.

Finally, the gCBT is largely insensitive to variations in the gate capacitance, as shown in the inset of Fig.~\ref{fig:mcmc}(b), where the charging energy becomes essentially independent of the gate or junction capacitance variations for a fixed $C_\Sigma$. This is in sharp contrast to the jCBT where $C_\Sigma\approx 2C_j$ and the $E_c$ curve becomes extremely steep and sensitive to variations in $C_j$. The insensitivity of the gCBT to $C_j$ variations widely eases the requirements on the nanofabrication process. Note also that the charging energy of the gCBT is only about half of the value of the jCBT at constant $C_\Sigma$. For the purpose of performing thermometry at the lowest temperatures, this is of course a valuable advantage, since it extends the range of validity of the thermometer, for a given $C_\Sigma$, to lower temperatures by a factor of 2: the range of the gCBT reaches down to temperatures of $\approx$~0.10$(e^2/C_\Sigma)/k_B$, whereas the jCBT is limited to $\approx$~0.80$(e^2/C_\Sigma)/k_B$, with elementary charge $e$. This is on top of the effect of the extra capacitance gain due to the top metal gate. The total capacitance $C_\Sigma$, rather than the charging energy, is the new relevant physical quantity to fix and to scale temperature with for a comparison e.g. of a gCBT and a jCBT, see Fig.~\ref{fig:mevsmcmc}, resulting in the same average conductance in the universal regime.


\subsection*{The Demagnetization Process}
\begin{figure}[ht]
	\centering
	\includegraphics[width=\linewidth]{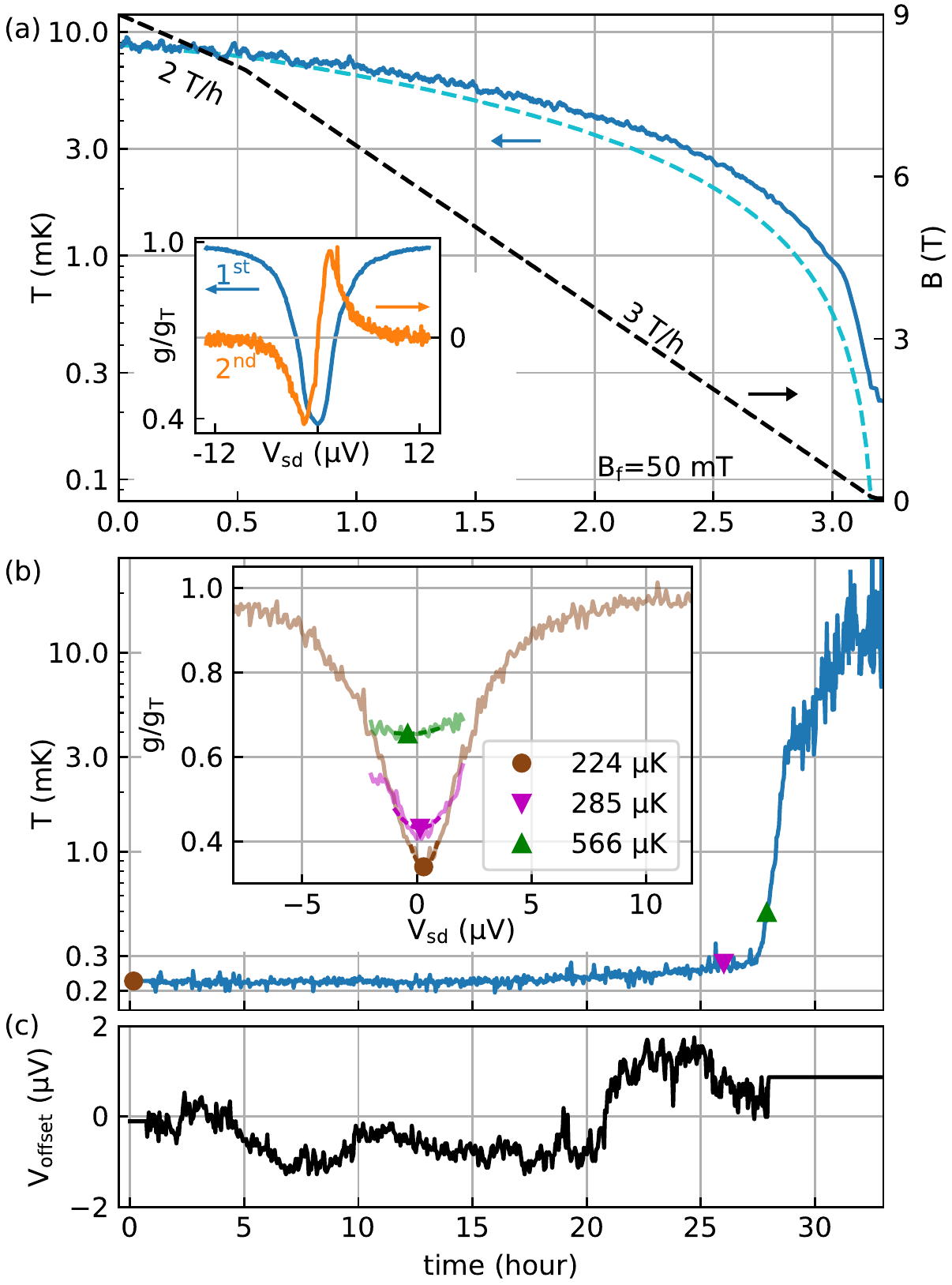}
	\caption{{\bf The demagnetization process} (a) 
		The magnetic field (black dashed line on the linear right axis) is ramped down, as indicated. The measured CBT temperature (dark blue) is compared with the ideal demagnetization (dashed light blue). Inset: below 500~mT, the second harmonic signal (orange) is used to keep the bias voltage at the center of the conductance dip (blue). (b) The warmup curve versus time on a logarithmic scale. Inset: bias traces at 3 different temperatures, as labelled. During the warmup, $V_\mathrm{sd}$ was scanned continuously and the offset voltage was determined from the minimum of a parabolic fit (dashed lines in the inset). The voltage drift over time is displayed in panel (c).
	}
	\label{fig:demag}
\end{figure}

First we magnetize the NRs in a large initial magnetic field $B_\mathrm{i}$. This creates a significant nuclear-spin polarization and releases the heat of magnetization which is absorbed over several days of precooling by the mixing chamber. After that, we ramp down the magnet to a final field $B_\mathrm{f}>0$ slowly to minimize eddy current heating and maintain adiabaticity. For an ideal demagnetization, the process is fully adiabatic, i.e. no heat enters or leaves the nuclear-spins subsystem. In this case, the temperature is reduced by the same factor as the magnetic field is ramped down, and the efficiency $\xi=(T_\mathrm{i}/T_\mathrm{f}) \div (B_\mathrm{i}/ B_\mathrm{f}) $ is equal to 1. In an experiment, however, there is always some heat leaking into the system, reducing the efficiency below 1. In Fig.~\ref{fig:demag}(a), we show the CBT temperature during demagnetization (dark blue curve), as well as the ideal demagnetization curve (light blue curve), both starting from the same initial temperature $T_\mathrm{i}=8.6$~mK. For the run shown here, the magnetic field was reduced from 9~T by a factor of 180 to 50~mT, while the CBT cools by a factor of 39, corresponding to an efficiency of about 22\%.

At the lowest temperatures, the CBT is very sensitive to the voltage bias, since the conductance dip, shown in the inset of Fig.~\ref{fig:demag}(b), is very narrow. Therefore, any voltage offsets, drifting over time, need to be carefully corrected continuously, as shown in panel (c) where the corrected offset voltage is plotted. During demagnetization, the second harmonic signal from the lock-in amplifier is monitored, shown in the inset of Fig.~\ref{fig:demag}(a), and $V_\mathrm{sd}$ is shifted to the left or right based on the sign of the second harmonic signal. At the lowest temperatures, the 2nd harmonic requires too large AC bias to be measurable. Instead, short scans of $g$ versus $V_\mathrm{sd}$ are taken over a small range; a parabolic fit, shown in the inset of Fig.~\ref{fig:demag}(b), determines the offset voltage which should be compensated. We note an extreme sensitivity to such offset voltages: about 2~µV corresponds to the extracted temperature being incorrectly large by a factor of 2.

The minimum conductance extracted from a parabolic fit around zero bias corresponds to 224$\pm$7~µK, reached within minutes after arriving at the final field. This time is still relatively fast even at these very low temperatures due to the very small heat capacity of electrons compared to the nuclear spins which provide the cooling. The uncertainty used here corresponds to the conservative error estimate as described above in Fig.~\ref{fig:mcmc}(c). Over time, the CBT slowly warms up, but stays below 300~µK for over 27 hours. The temperature then rises quickly to the temperature of the mixing chamber and above. 

\subsection*{The Heat Leaks}
\begin{figure}[ht]
	\centering
	\includegraphics[width=\linewidth]{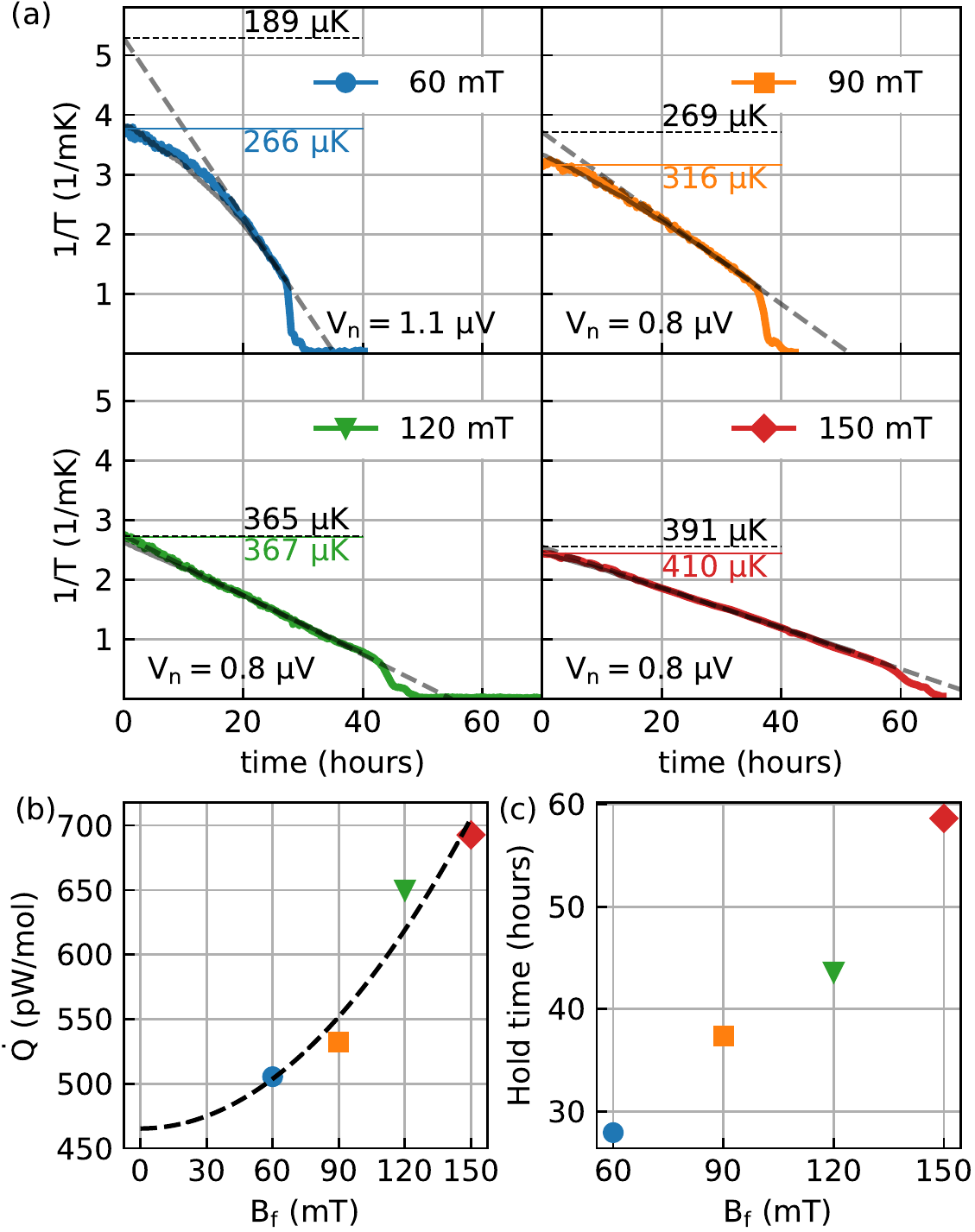}
	\caption{{\bf Demagnetization to different final fields} (a) The warmup curves for 4 different final fields $B_\mathrm{f}$ on a 1/T scale. The dashed lines are fit to the linear part at the end of each warmup. The solid black lines are the voltage-noise-corrected reading, assuming that the dashed line is the true temperature of electrons. (b) The heat leaks per mole of copper into each island extracted from slopes of dashed lines in panel (a). The fit to the parabola suggests that there is 457 pW/mol of field-independent heat leaks into the islands. (c) The time between the end of demagnetization and when the temperatures rises above 2~mK.}
	\label{fig:heatleaks}
\end{figure}

We repeated the demagnetization cycle for various final fields, shown in Fig.~\ref{fig:heatleaks}(a), to gain insight into the nature of the heat leaks. If a constant heat leak is present, the plot of the inverse temperature as a function of time turns out to be linear, and the slope can be used to extract the heat leak \cite{Pickett1988AND, Pobell2007AND}. For both $B_\mathrm{f}$=120~mT and 150~mT, the curves are indeed linear for up to 40 or 50 hours, when finally a rapid warmup occurs. For the two lower final fields, however, the warmup curve is initially nonlinear, allowing a good linear fit only  several hours after the end of the ramp down. In the nonlinear regime, the CBT consistently reads a higher temperature compared to the back-extrapolated linear fit (black dashed lines). This may be attributed to voltage noise $V_\mathrm{n}$ across the CBT leads. Due to the very narrow conductance dip at the lowest temperatures, voltage noise effectively averages over the center of the conductance dip, and artificially increases $g_0$ readings which in turn produce higher apparent CBT temperatures. This can be calculated by smearing the MCMC model of $g(V_\mathrm{sd})$ with gaussian noise. Choosing noise values around $V_\mathrm{n}$=1~µVrms, as labelled, makes the model (grey curves) almost entirely coincide with the measured temperatures. Given the good agreement, this makes the extrapolated temperature of 189~µK quite plausible as the actual electron temperature when the CBT reads 265~µK at the beginning of the warmup at $B_\mathrm{f}$=60 mT.

The heat leaks extracted from the warmup slopes for various final fields are shown in Fig.~\ref{fig:heatleaks}(b) and are as low as $\approx$0.5~nW per mole of NR material on the CBT. This corresponds to 2.8~aW/island, significantly lower than previous warm-up experiments \cite{Sarsby2020, Bradley2016}. Given the low heat leaks, the electron temperature rises only about 5\% above the nuclear spin temperature even at the lowest temperatures shown here. A large part of this heat leak, about 0.47~nW/mol, is magnetic-field independent and is probably due to heat releases from small amounts of epoxy, glue, varnish, paper, or plastic, which are used for constructing various parts of the setup and sample holder in particular. In addition, there is a smaller component that scales quadratically with $B_\mathrm{f}$ (black dashed parabola). This field dependence of eddy-current heating is expected due to mechanical vibrations in an inhomogeneous magnetic field.

We note that close to the end of every warmup in Fig.~\ref{fig:heatleaks}(a), above about 1~mK, there is a rapid rise of temperature, more quickly than linear in $1/T$ from the on-chip NR warmup. First, the CBT heats up rapidly to a plateau around 5~mK, also visible in Fig.~\ref{fig:demag}(b), and then warms up to even higher temperatures. This can be explained by the sequential warmup of the bond pads and the sample box. See the supplementary materials for details. Eventually, after the nuclear heat capacities are exhausted, cooling can occur only through the superconducting heat switches, providing a very weak link to the mixing chamber, thus warming up the CBT well above $T_\mathrm{MC}$, typically $\approx$30~mK.

Finally in Fig.~\ref{fig:heatleaks}(c), we plot the hold time, defined here as the time-span between the end of demagnetization and when the temperature of the CBT exceeds 2~mK. Despite slightly larger heat leaks at higher final fields, the hold times are longer for larger fields due to the enhanced nuclear heat capacity which scales as $B_\mathrm{f}^2$.

\section{Conclusion and Outlook}
We reduce the temperature of on-chip electrons to 224$\pm$7~µK. Heat leaks that are a fraction of a nanowatt per mole of NR material allow the temperature to stay below 300~µK for over 27 hours. These low heat leaks are significant since the experiment was mounted on a pulse tube cryogen-free dilution refrigerator which, while offering many advantages such as large experimental space and low cost of operation, induce mechanical vibrations at multiple harmonics of 1.4~Hz. This can be attributed to a carefully designed sample box that surrounds the sample with NR material and provides a microkelvin environment, non-inductive microwave filters, mechanically rigid wiring inside the demagnetization field, and blocking external heat leaks using macroscopic plates of NR materials on individual leads.

A new type of CBT, termed gCBT, is introduced. It has a metal top gate that covers all islands, resulting in a dominant island-to-ground capacitance and a significantly enhanced total island capacitance. The gCBT used here has a very low charging energy of 737~µK$\times k_B$ which allows temperature measurements deep into the microkelvin range. The gCBT also offers improved accuracy despite random charge offsets on islands. At $0.2E_c\approx160$~µK the thermometer has an accuracy of better than 10\%.

The current limitations are a saturated temperature reading due to voltage noise and a minimum final field to suppress superconductivity. In perpendicular field configuration, the final field could be reduced by another factor of about 3-4. Further, scaling to longer chains (or improved filtering) could reduce the voltage noise, and thus improve the temperature reading of the CBT. The range of validity could also be further extended by increasing the total capacitance with optimizations such as a thinner oxide, an improved dielectric or by adding a bottom gate. Such improvements would potentially make temperatures below 50~µK accessible. The gCBT is in-principle easy to fabricate and could be integrated on-chip with semiconductors and other, exotic materials as both a thermometer and a cooler, thus opening new avenues for electronic transport experiments at ultra-low temperatures.

\section*{Materials and methods}
The fabrication process of the gCBT starts with an undoped silicon substrate, with a 300~nm thermally-grown SiO$_2$ layer covering its surface. The tunnel junctions are made of a layer of SiO$_2$, sandwiched between two layers of aluminum. They are fabricated \emph{ex situ}, allowing for control over the capacitance and resistance of the junctions independently \cite{Prunnila2010ExSituCBT}. On each island 39~µm$\times$206~µm$\times$5~µm of copper is electroplated. This acts as the \emph{on-chip} NR. The islands are then covered with 400~nm of a dielectric using atomic layer deposition (ALD), followed by a layer of TiW and 200~nm of gold forming a metallic gate on top of islands (see Fig.~\ref{fig:cbtbox}(a-c)). This gate is permanently grounded to the body of the copper box in this experiment. From the value of $g_T$=21.68~µS, we estimate the average junction resistance to be $1/g_T \times M/N$=28~k$\Omega$.


All measurements are performed using a 2-point voltage-bias current-measurement scheme. A standard lock-in amplifier measures the differential conductance $dI/dV$ at 13~Hz with an AC excitation between 0.1-2~µVrms. Conductance $g$ versus $V_\mathrm{sd}$ sweeps are performed  slowly to allow for the steady-state of the heat flow to be reached. A current-to-voltage converter\cite{BasPI} set to $10^8$ V/A with a bandwidth of 3~kHz amplifies the signal before it is fed to the lock-in, and it is feedback stabilized to less than 100~nV/$^\circ$C. A high precision digital-to-analogue converter\cite{BasPI} controls the DC source-drain bias voltage.

The master equation used to extract temperature from conductance is approximated at zero-bias voltage by the following 3\textsuperscript{rd} order polynomial:
\begin{align}
	\delta g = \frac{u}{6} - \frac{u^2}{60} + \frac{u^3}{360}
\end{align}
where $u=2E_c/k_B T$. Note that in the definition of $u$, there is a factor of 2, consistent with the original definition by Pekola and coworkers \cite{Pekola1994}. This factor was later dropped by some authors \cite{Meschke2004,Feshchenko2015,Palma2017OnOff} which lead to a different definition of the charging energy.

For the demagnetization processes shown here, $B_\mathrm{i}$ is always 9~T. Precooling times vary between 2 days for data shown in Fig.~\ref{fig:heatleaks} and 5 days for Fig.~\ref{fig:demag}. The latter is the demagnetization with the longest precooling time (see Fig.~\ref{fig:precool}) and the lowest final temperature. The field is ramped down at 2~T/hr for B>8~T, and 3~T/hr otherwise. The lower rate is used to avoid magnet quenches at high field. $B_\mathrm{f}$ is limited to $\approx$45~mT, the minimum field required to prevent the superconducting transition of aluminum films across the tunnel junctions in the CBT.


During demagnetization, when the field drops below 500~mT, we use the second harmonic signal from the lock-in amplifier to keep the DC voltage close to the minimum of the conductance dip $V_0$. The second harmonic signal, the orange curve in the inset of Fig.~\ref{fig:demag}(a), crosses zero sharply at $V_0$, hence its sign can be used to adjust the DC voltage offset. During the warm up, we reduce the AC excitation to 100~nV which makes the second harmonic signal undetectable. We therefore use $V_\mathrm{sd}$ scans in a narrow range between -2 and 2~µV and a parabolic fit to find $V_0$, shown in the inset of Fig~\ref{fig:demag}(b). This range is narrow enough to prevent significant Joule heating, and wide enough to capture the conductance minimum.


The copper box shown in Fig~\ref{fig:cbtbox}(d) is rigidly attached to the body of the fridge, and its body, made of oxygen-free high-conductivity (OFHC) copper and electroplated with gold, acts as a radiation shield as well as an NR, enclosing the sample in a microkelvin environment towards the end of demagnetization. The silver-epoxy microwave filters \cite{Scheller2014SilverEpoxy} are two sets of coils that are wound in opposite directions, or non-inductively, and placed perpendicular to the direction of the magnetic field. Each signal wire is filtered separately and then attached to a bonding pad which is an L-shaped slab of gold-plated copper, placed inside the cavity and gold-bonded to the sample. The pads act as extra NRs inside the closure and after the microwave filters.

The warm-up time, the time it takes for the temperature of the nuclear spins to rise from $T_{n,1}$ to $T_{n,2}$, is expected\cite{Pobell2007AND} to be
\begin{align}
	t=\left( \frac{\lambda_n B_\mathrm{f}^2}{\mu_0 \dot{Q}}\right)\left(\frac{1}{T_{n,1}} - \frac{1}{T_{n,2}}\right)
\end{align}
where $\dot{Q}$ is the molar heat leak into the on-chip NRs, $\lambda_n$ is the Curie constant, $\mu_0$ is the vacuum permeability. Assuming all the heat leak from the electronic subsystem to enter the nuclear-spins subsystem through the Korringa link, the electron and nuclear spin temperature are related via
\begin{align}
	\frac{T_e}{T_n}=1+\left( \frac{\mu_0 \kappa \dot{Q}}{\lambda_n B_\mathrm{f}^2}\right)
\end{align}
Here $\kappa$ is the Korringa constant. We can therefore relate the derivative of inverse temperature to the heat leak
\begin{align}
	-\frac{\partial}{\partial t}\left(\frac{1}{T_e}\right)=\left(\frac{\lambda_n B_\mathrm{f}^2}{\mu_0 \dot{Q}}+\kappa\right)^{-1}
\end{align}
Note that for constant $\dot{Q}$, the $1/T_e$ is linear in time and the slope of the line can be used to extract the constant heat leak $\dot{Q}$.


\begin{thebibliography}{10}

\bibitem{Hasan2010TopologicalOrdering}
M.~Z. Hasan, C.~L. Kane, {\it Rev. Mod. Phys.\/} {\bf 82}, 3045 (2010).

\bibitem{Braunecker2013TopologicalMajorana}
B.~Braunecker, P.~Simon, {\it Phys. Rev. Lett.\/} {\bf 111}, 147202 (2013).

\bibitem{Chekhovich2013}
E.~A. Chekhovich, {\it et~al.\/}, {\it Nature Materials\/} {\bf 12}, 494
  (2013).

\bibitem{Chesi2008QHFerromagnet}
S.~Chesi, D.~Loss, {\it Phys. Rev. Lett.\/} {\bf 101}, 146803 (2008).

\bibitem{Mackenzie2003Sr2RuO4}
A.~P. Mackenzie, Y.~Maeno, {\it Rev. Mod. Phys.\/} {\bf 75}, 657 (2003).

\bibitem{Cooper2009}
N.~R. Cooper, A.~Stern, {\it Phys. Rev. Lett.\/} {\bf 102}, 176807 (2009).

\bibitem{Kapitulnik2019DisorderedSuperconductor}
A.~Kapitulnik, S.~A. Kivelson, B.~Spivak, {\it Rev. Mod. Phys.\/} {\bf 91},
  011002 (2019).

\bibitem{Hanson2007}
R.~Hanson, L.~P. Kouwenhoven, J.~R. Petta, S.~Tarucha, L.~M.~K. Vandersypen,
  {\it Rev. Mod. Phys.\/} {\bf 79}, 1217 (2007).

\bibitem{Scheller2014SilverEpoxy}
C.~P. Scheller, {\it et~al.\/}, {\it Applied Physics Letters\/} {\bf 104},
  211106 (2014).

\bibitem{Pickett1988AND}
G.~R. Pickett, {\it Reports on Progress in Physics\/} {\bf 51}, 1295 (1988).

\bibitem{Pobell2007AND}
F.~Pobell, {\it Matter and Methods at Low Temperatures\/} (Springer-Verlag
  Berlin Heidelberg, 2007).

\bibitem{Knuuttila2001}
T.~A. Knuuttila, {\it et~al.\/}, {\it Journal of Low Temperature Physics\/}
  {\bf 123}, 65 (2001).

\bibitem{Bradley2016}
D.~I. Bradley, {\it et~al.\/}, {\it Nature Communications\/} {\bf 7}, 10455
  (2016).

\bibitem{Iftikhar2016}
Z.~Iftikhar, {\it et~al.\/}, {\it Nature Communications\/} {\bf 7}, 12908
  (2016).

\bibitem{Palma2017MagneticCooling}
M.~Palma, {\it et~al.\/}, {\it Review of Scientific Instruments\/} {\bf 88},
  043902 (2017).

\bibitem{Saira2012}
O.-P. Saira, A.~Kemppinen, V.~F. Maisi, J.~P. Pekola, {\it Phys. Rev. B\/} {\bf
  85}, 012504 (2012).

\bibitem{Zorin1995}
A.~B. Zorin, {\it Review of Scientific Instruments\/} {\bf 66}, 4296 (1995).

\bibitem{Maradan2014}
D.~Maradan, {\it et~al.\/}, {\it Journal of Low Temperature Physics\/} {\bf
  175}, 784 (2014).

\bibitem{Ciccarelli2016}
C.~Ciccarelli, R.~P. Campion, B.~L. Gallagher, A.~J. Ferguson, {\it Applied
  Physics Letters\/} {\bf 108}, 053103 (2016).

\bibitem{Bradley2017OnChip}
D.~I. Bradley, {\it et~al.\/}, {\it Scientific Reports\/} {\bf 7}, 45566
  (2017).

\bibitem{Clark2010}
A.~C. Clark, K.~K. Schwarzwälder, T.~Bandi, D.~Maradan, D.~M. Zumbühl, {\it
  Review of Scientific Instruments\/} {\bf 81}, 103904 (2010).

\bibitem{Palma2017OnOff}
M.~Palma, {\it et~al.\/}, {\it Applied Physics Letters\/} {\bf 111}, 253105
  (2017).

\bibitem{Casparis2012}
L.~Casparis, {\it et~al.\/}, {\it Review of Scientific Instruments\/} {\bf 83},
  083903 (2012).

\bibitem{Feshchenko2015}
A.~V. Feshchenko, {\it et~al.\/}, {\it Phys. Rev. Applied\/} {\bf 4}, 034001
  (2015).

\bibitem{Jones2020}
A.~T. Jones, {\it et~al.\/}, {\it Journal of Low Temperature Physics\/}
  (2020).

\bibitem{Sarsby2020}
M.~Sarsby, N.~Yurttag{\"u}l, A.~Geresdi, {\it Nature Communications\/} {\bf
  11}, 1492 (2020).

\bibitem{Pekola1994}
J.~P. Pekola, K.~P. Hirvi, J.~P. Kauppinen, M.~A. Paalanen, {\it Phys. Rev.
  Lett.\/} {\bf 73}, 2903 (1994).

\bibitem{Meschke2004}
M.~Meschke, J.~P. Pekola, F.~Gay, R.~E. Rapp, H.~Godfrin, {\it Journal of Low
  Temperature Physics\/} {\bf 134}, 1119 (2004).

\bibitem{Feshchenko2013}
A.~V. Feshchenko, {\it et~al.\/}, {\it Journal of Low Temperature Physics\/}
  {\bf 173}, 36 (2013).

\bibitem{Farhangfar1997}
S.~Farhangfar, {\it et~al.\/}, {\it Journal of Low Temperature Physics\/} {\bf
  108}, 191 (1997).

\bibitem{Yurttagul2021}
N.~Yurttag{\"u}l, M.~Sarsby, A.~Geresdi, {\it Journal of Low Temperature
  Physics\/}  (2021).

\bibitem{Prunnila2010ExSituCBT}
M.~Prunnila, {\it et~al.\/}, {\it Journal of Vacuum Science \& Technology B\/}
  {\bf 28}, 1026 (2010).

\bibitem{BasPI}
{Basel Precision Instruments}, {Low-Noise High-Stability I-to-V Converter
  IF3602, and LNHR DAC voltage source 24 bit} (2019).

\bibitem{Hirvi1996}
K.~P. Hirvi, M.~A. Paalanen, J.~P. Pekola, {\it Journal of Applied Physics\/}
  {\bf 80}, 256 (1996).

\end{thebibliography}


\section*{Acknowledgements}
This research was supported by the EU H2020 European Microkelvin Platform (EMP) grant number 824109, innovation program under grant agreement number 766853 Energy Filtering Non-Equilibrium Devices (EFINED), by the Academy of Finland through the Centre of Excellence program numbers 336817 and 312294, the Swiss National Science Foundation grant number 179024, the Swiss Nanoscience Institute, and the Georg H. Endress Foundation.



\section*{Data availability}
Data generated for this study, along with the procedures for producing the figures, are available at \href{https://github.com/mohammadsamani/microkelvin}{https://github.com/mohammadsamani/microkelvin}.


\section{Supplementary Material}

\subsection{Markov-Chain Monte-Carlo Model for the CBT}
Coulomb blockade thermometry is implemented by tunnel junction arrays which are designed to operate within the boundaries of the orthodox theory of tunnelling. Numerical single-electronics is utilized to model the thermal response of the device. In the universal regime of charging a fast single-electron transistor (SET), the master equation approach is used to calculate the tunnel conductance at a given temperature. The charging energy is then only a universal scaling factor of the charging curve. When the universal regime of charging is left the tunnel conductance of a CBT can not be assumed to scale in a universal way, but is determined by the exact charging physics of the tunnel junction array and offset charge on the islands \cite{Yurttagul2021}. Depending on the details of the electrostatic coupling of the islands by their mutual capacitance and their capacitance to ground, the conductance has a characteristic scaling with $T$.

\begin{figure}[ht]
	\centering
	\includegraphics[width=\linewidth]{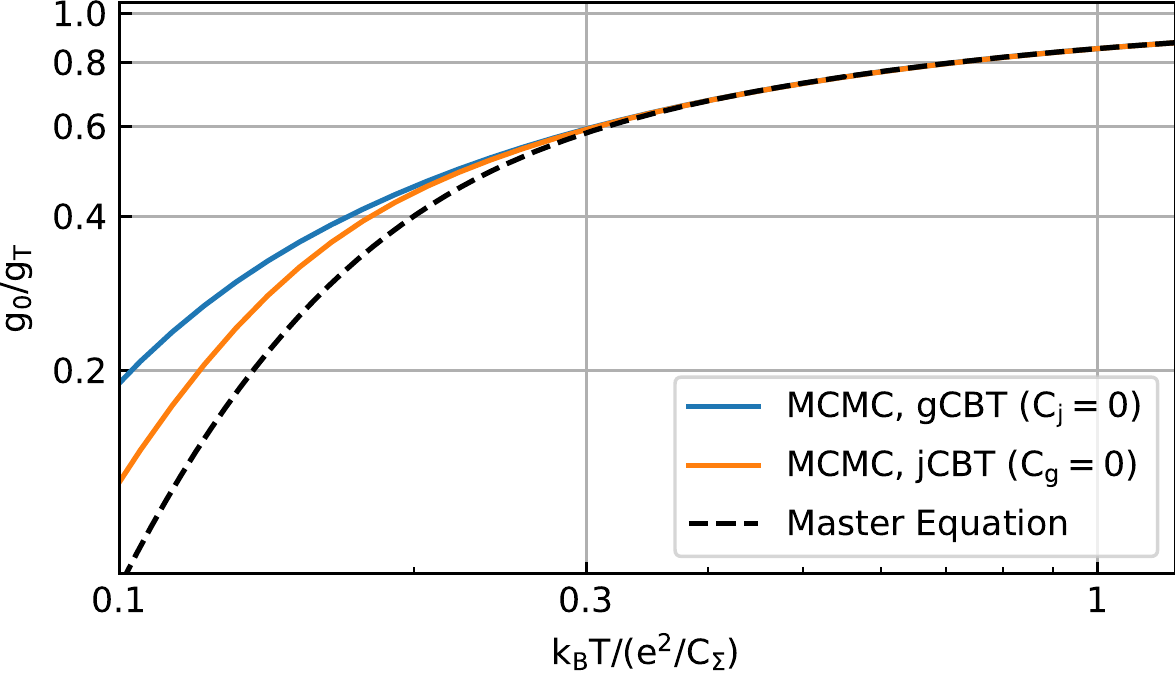}
	\caption{{\bf Comparing tunnel conductance models} Tunnel conductance $g/g_T$ calibration of the gCBT plotted versus dimensionless temperature $1/u=k_B T/(e^2/C_\Sigma)$, calculated with the master equation model and the Monte-Carlo model (MCMC) for a junction array with $N=33$ junctions and a given $C_\Sigma=2C_j+C_g$. The MCMC curves are calculated with random offset charges $C_\Sigma$ is the same in both. For the orange curve $C_\Sigma=2C_j$, while for the blue curve $C_\Sigma=C_g$.}
	\label{fig:mevsmcmc}
\end{figure}

For calculating the tunnel conductance of the CBT we utilize a Markov-Chain Monte-Carlo (MCMC) algorithm which was introduced in detail in Ref. \citenum{Hirvi1996}. We assume a uniform array in which the island capacitance $C_{\Sigma i}=2C_{ji}+C_{gi}$ consists of the mutual junction capacitance $C_{ji}=C_j$ and the capacitance to ground $C_{gi}=C_{g}$. Note that $j$ in $C_{ji}$ is not an index and indicates \emph{junction} capacitance. Previously the MCMC model was only utilized in the limits of $C_j\gg C_g$, but here we extend the model to the gCBT with $C_\mathrm{g} \geq C_j$. Charge conservation is given by
\begin{align}
	C_j\varphi_{i-1}-\left(2C_j+C_g\right)\varphi_i+C_j\varphi_{i+1}=-q_i
	\label{cmatrix}
\end{align}
Here, $q_i=\pm ne+q_o$ is the total island charge, which is the sum of the integer number of electrons or holes on the island plus the continuous offset charge $q_o$. The electrostatic free energy of the array is determined by the electrostatic island potentials $\varphi_i$ according to
\begin{align}
	F_{el}=\frac{1}{2}\sum_{i=1}^{N}C_{ji}\left(\varphi_{i}-\varphi_{i-1}\right)^2+\frac{1}{2}\sum_{i=1}^{N-1}C_{gi}\varphi_{i}^2
	\label{el_energy}
\end{align}
The tunnelling rates are calculated in the limits of the orthodox theory of tunneling, assuming that the junctions with resistance $R_i$ are sufficiently opaque to suppress cotunnelling over the array and the free energy change of the system is only due to charging and not due to interaction with the electromagnetic environment. The latter is usually well established by the high impedance environment of the islands in a tunnel junction array. The tunnelling rate of charges over the i\textsuperscript{th} junction with resistance $R_i$ are
\begin{align}
	\Gamma_i^\pm=\frac{1}{e^2R_i}\frac{\Delta F_i^\pm}{1-\textrm{exp}(-\Delta F_i^\pm/k_\textrm{B}T)}.
\end{align}
Here $\Gamma^+$ implies forward (with respect to the bias voltage $V_\mathrm{sd}$) tunnelling and $\Gamma^-$ backwards tunnelling over the i\textsuperscript{th} junction with $\Delta F_i^\pm = \Delta F_{el} \mp e V_\text{sd} \cdot R_i/R_\Sigma$. The tunnelling time and transferred charge is calculated by stochastic sampling \cite{Hirvi1996} of single tunnelling events over all junctions as
\begin{align}
	\Delta t_p=\left(\sum_{i=1}^N \Gamma_i^++\Gamma_i^-\right)^{-1},
\end{align}
\begin{align}
	\Delta Q_p=\displaystyle e\frac{\sum_{i}\left(\Gamma_i^+-\Gamma_i^-\right)R_i/R_\Sigma}{\sum_{i}\left(\Gamma_i^++\Gamma_i^-\right)}
\end{align}
The index $p$ relates each $\Delta t$ and $\Delta Q$ to a specific Markov-chain element. The total tunnel current is $I=\sum_p\Delta Q_p/\sum_p\Delta t_p$ and the differential conductance is $g=\text{d}I/\text{d}V_\text{sd}$.

A specific choice of background charge $q_o$ makes each Markov-chain unique which is why random background charge is included by calculating the conductance distribution $P(g)$ with a sufficiently large number of random configurations of $q_o$, which are themselves chosen either as fully random or physical offset charge. Fully random offset charges are generated from a uniform distribution with a large width of 100e around zero. For calculating physical offset charges a fully random offset charge configuration is altered by single electron tunnelling until the electrostatic free energy reaches a minimum (all $\Delta F_i^\pm$ are positive).

The universal charging regime is usually defined for the local charging limit, or jCBT, only in which tunnel junction arrays of length $N$ have the same tunnel conductance at the same $k_B T/E_c$ with $E_c=(N-1)/N\times e^2/C_\Sigma$. However the universal regime holds also further for CBT with the same $N$ and $C_\Sigma$, but a different $E_c$. The exact universal scaling factors can only be determined with a MCMC approach.
For demonstrating this we calculate the tunnel conductance of tunnel-junction arrays which have an equal value of $C_\Sigma=2C_j+C_g$ but are either in the non-local limit ($C_j=0$) or in the local limit ($C_g=0$). The result is plotted in Fig.~\ref{fig:mevsmcmc}, where the normalized tunnel conductance $g_0/g_T$ is plotted against the dimensionless factor $1/u=k_\text{B}T/(e^2/C_\Sigma)$. As can be seen, arrays with equal $C_\Sigma$ have an identical charging behaviour in the universal regime which makes $C_\Sigma$ the right scaling factor and not the charging energy $E_c$. Remarkably, $E_c$ differs by nearly a factor of two for the blue and the orange curves in Fig.~\ref{fig:mevsmcmc} which establishes the enhanced immunity to offset charges of CBTs with high $C_g$.

\subsection{The voltage-noise model}
Since the width of the conductance dip at very low temperatures is comparable to the magnitude of the voltage noise on measurement lines, the voltage noise smears $V_\mathrm{sd}$ around zero faster than the voltage-drift correction mechanism, which normally takes several minutes.

\begin{figure}[ht]
	\centering
	\includegraphics[width=\linewidth]{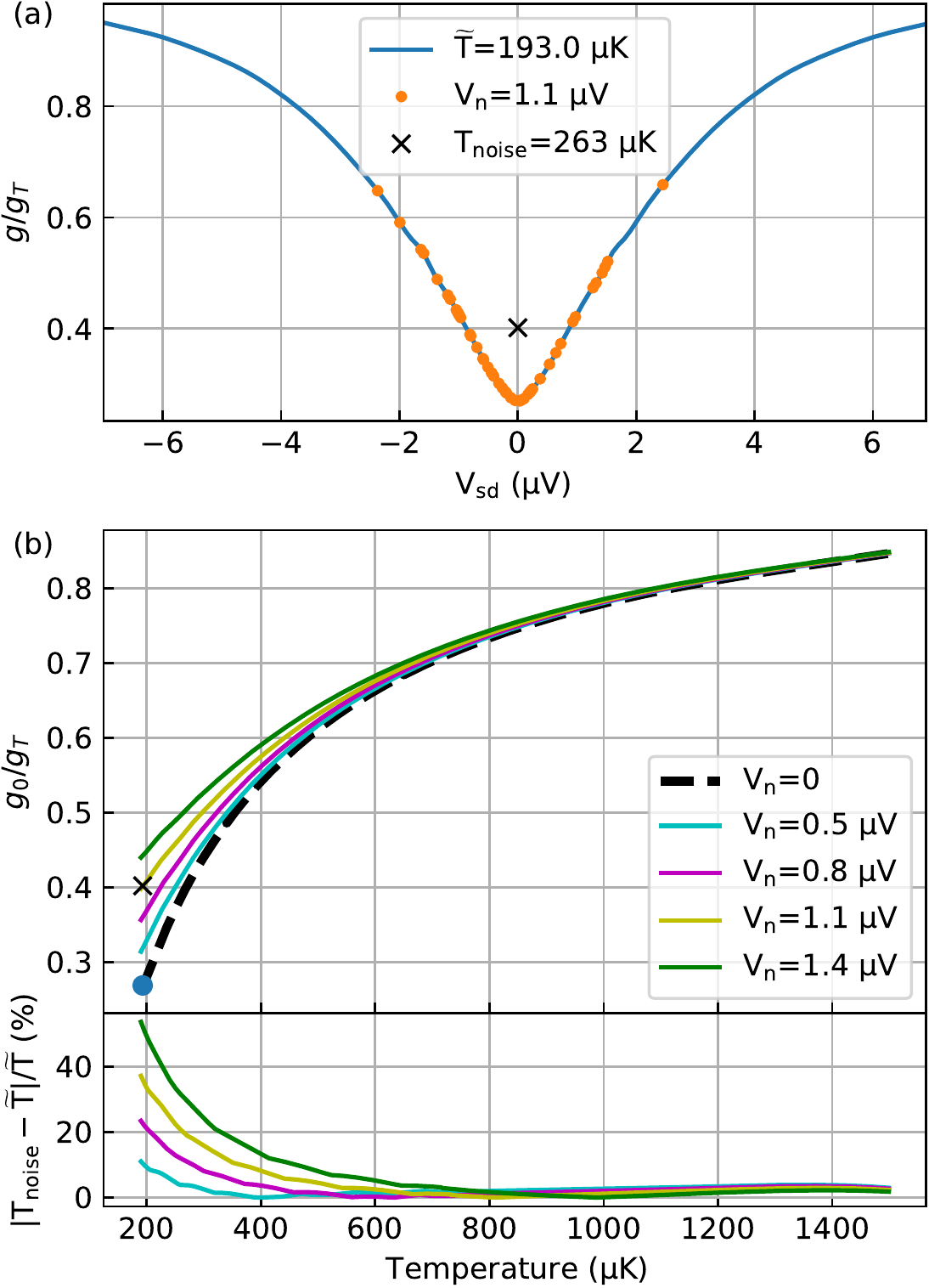}
	\caption{{\bf Voltage noise model} (a) The blue curve is the MCMC model for the ``true'' temperature $\widetilde{T}$=193~µK. Normally distributed random numbers with mean zero and standard deviation $V_\mathrm{n}$=1.1~µV are generated to represent voltage noise, orange dots, and their corresponding conductance are calculated and averaged, resulting in the black cross. From this number, $T_{noise}$ is obtained. (b) The expected relative conductance is shown for various values of $V_\mathrm{n}$ in the upper panel, and the relative error caused by voltage noise is shown in the lower panel.}
	\label{fig:voltagenoise}
\end{figure}

To model this effect, we generate normally distributed random noise on $V_\mathrm{sd}$ (orange dots in Fig.~\ref{fig:voltagenoise}) with mean 0~V and standard deviation $V_\mathrm{n}$. We then calculate the normalized conductance $g/g_T$ for each point based on a conductance curve (blue curve) at a given ``true'' temperature $\widetilde{T}$, and find the average conductance (black cross). This number is taken to be the \emph{noised} conductance in the model and is used to calculate a noised temperature. If the true temperature is 193~µK, for example, the noised temperature is expected to be 262~µK. In Fig.~\ref{fig:voltagenoise}, 40 points are shown for clarity, but $10^4$ points are used in the simulation to achieve a stable convergence.

At the lowest temperatures relevant to this work, the effect of the voltage noise can create an overestimated temperature reading by $\approx$40\%.

In Fig.~\ref{fig:heatleaks}(a), we use this technique to calculate the noised temperature (solid black curve), taking the back-extrapolated line (dashed) as the ``true'' temperature.

\subsection{Pulse tube vibrations}
Previously, clear evidence of the effect of mechanical vibrations driven by the pulse tube was demonstrated as voltage noise on measurement lines \cite{Palma2017OnOff}. The first approach to resolving this issue in the current experiment was to turn the pulse tube off during the most sensitive period of the demagnetization, B<2~T, when the heat capacity of the nuclear-spins system ($\propto B^2$) is low, but the heat leaks into the electronic system, due to vibrating conductors in an inhomogeneous magnetic field, are not negligible. This was especially a problem with the old sample box in which many contacts were made with thin, long, dangling wires that the pulse tube vibrations could easily shake.

\begin{figure}[ht]
	\centering
	\includegraphics[width=\linewidth]{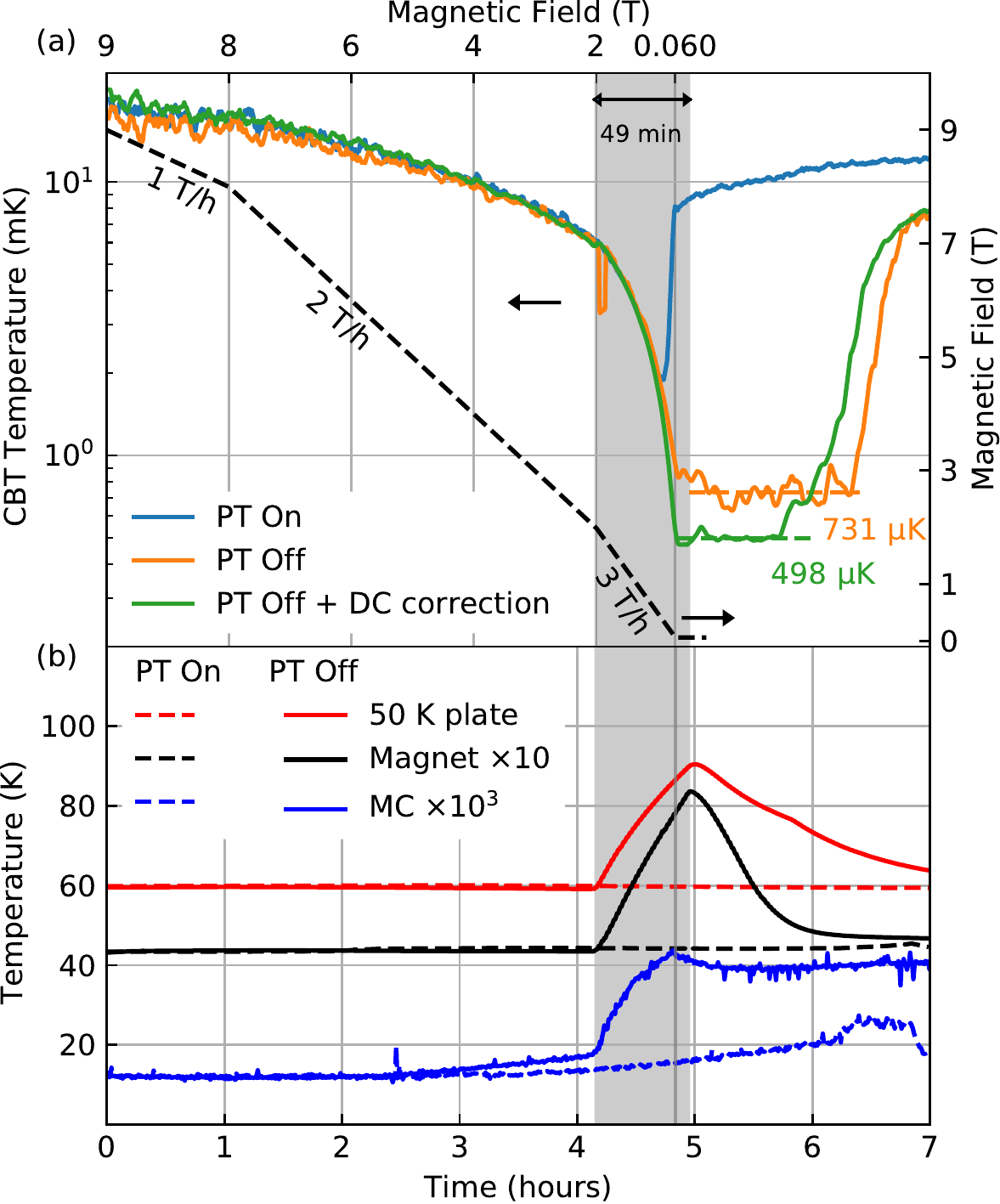}
	\caption{{\bf Turning the pulse tube off} (a) Three demagnetization attempts. The pulse tube stays on at all times for the blue curve, but for the orange and green curves, the pulse tube is turned off for about 49 minutes over the shaded area. After the final field is reached, the DC bias is stabilized for the green curve, but not for the orange curve. (b) The temperatures of various flanges of the dilution refrigerator rise during the time the pulse tube is off (solid). Compare to the stable temperatures (dashed) when the pulse tube always stays on.}
	\label{fig:pulsetubeonandoff}
\end{figure}

In Fig.~\ref{fig:pulsetubeonandoff}(a), 3 demagnetization attempts with similar magnetic field profiles are shown. In the first attempt (blue curve), the pulse tube stays on. We observe that the minimum temperature is just above 2~mK and there is essentially no hold time. Next we turn the pulse tube off (orange curve) from 2~T down to the final field (shaded grey area). The temperature plunges below 1~mK and stays there for about an hour. If the DC voltage is not stabilized (orange), the voltage drift in $V_\mathrm{sd}$ makes it seem like the temperature is fluctuating. DC stabilization corrects the temperature reading (green curve). The marginally lower final temperature is presumably due to a longer precooling time and slightly different demagnetization conditions.

\begin{figure}[ht]
	\centering
	\includegraphics[width=\linewidth]{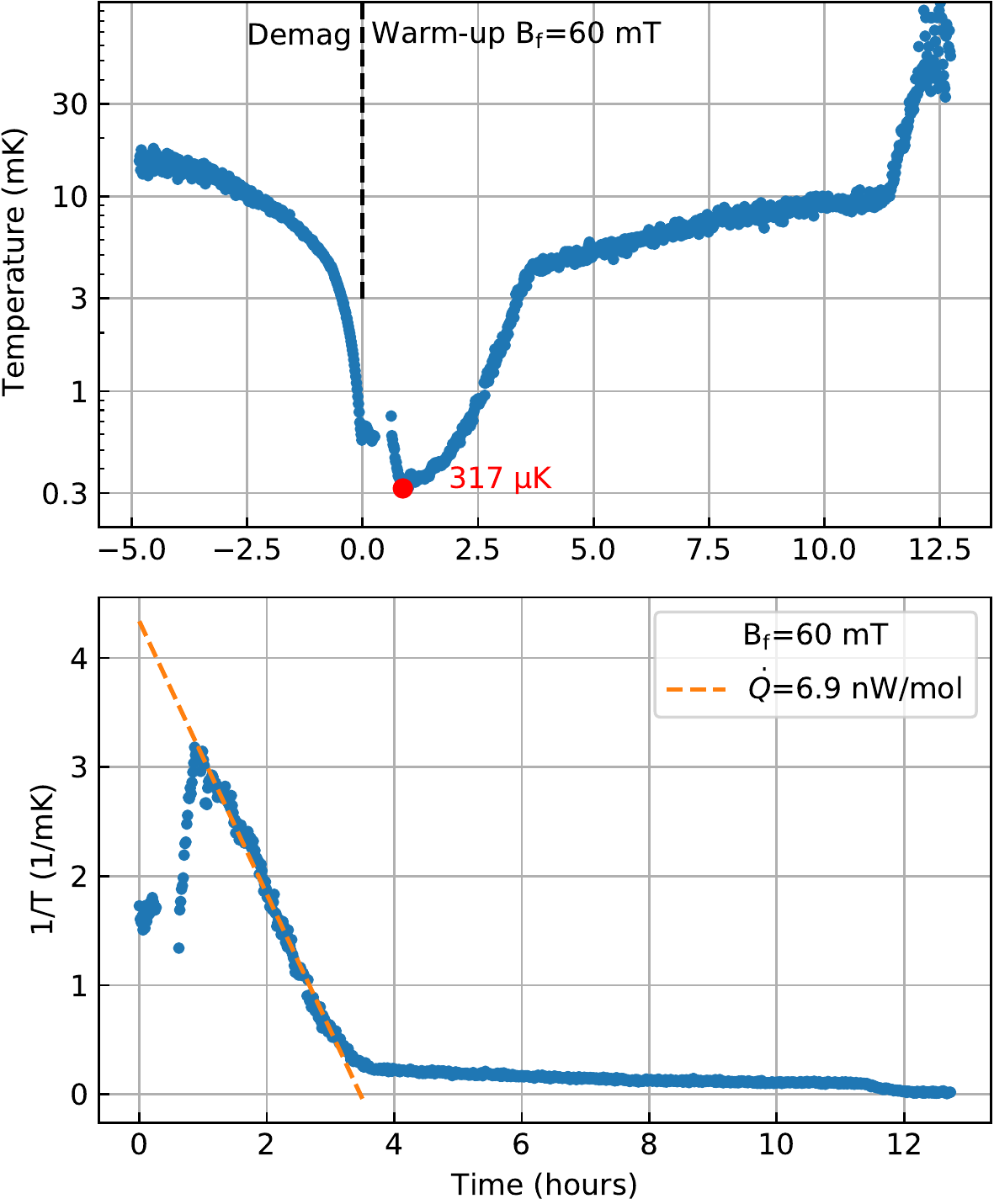}
	\caption{{\bf Warmup without interruption} Top: Demagnetization and warmup curves in a previous cooldown. Bottom: the warmup curve on a 1/T scale, showing an uninterrupted rise in temperature of the CBT to high temperatures, indicating that the CBT islands warm up before other components in the fridge.}
	\label{fig:uninterruptedwarmup}
\end{figure}

Of course turning off the pulse tube causes various parts of the fridge to warm up, as shown in Fig.~\ref{fig:pulsetubeonandoff}(b). We turn the pulse tube back on after about 49 minutes, when the magnet temperature approaches 9~K. 

For longer hold times below 1~mK and more stable measurement conditions, a more robust mechanical design of the sample box and the wires inside the magnetic field were needed.

Before warming up the fridge to room temperature, a demagnetization with a very long precooling (>400 hours) was performed, shown in Fig.~\ref{fig:uninterruptedwarmup}. We note that for the initial part of the warmup a linear in $1/T$ behaviour is observed, as expected for the thermodynamic warmup of the NR material on the CBT island under a constant heat leak. A heat leak of 6.9~nW/mol or 32~aW/island is extracted, similar in size to previous works \cite{Palma2017OnOff}.

\begin{figure}[ht]
	\centering
	\includegraphics[width=\linewidth]{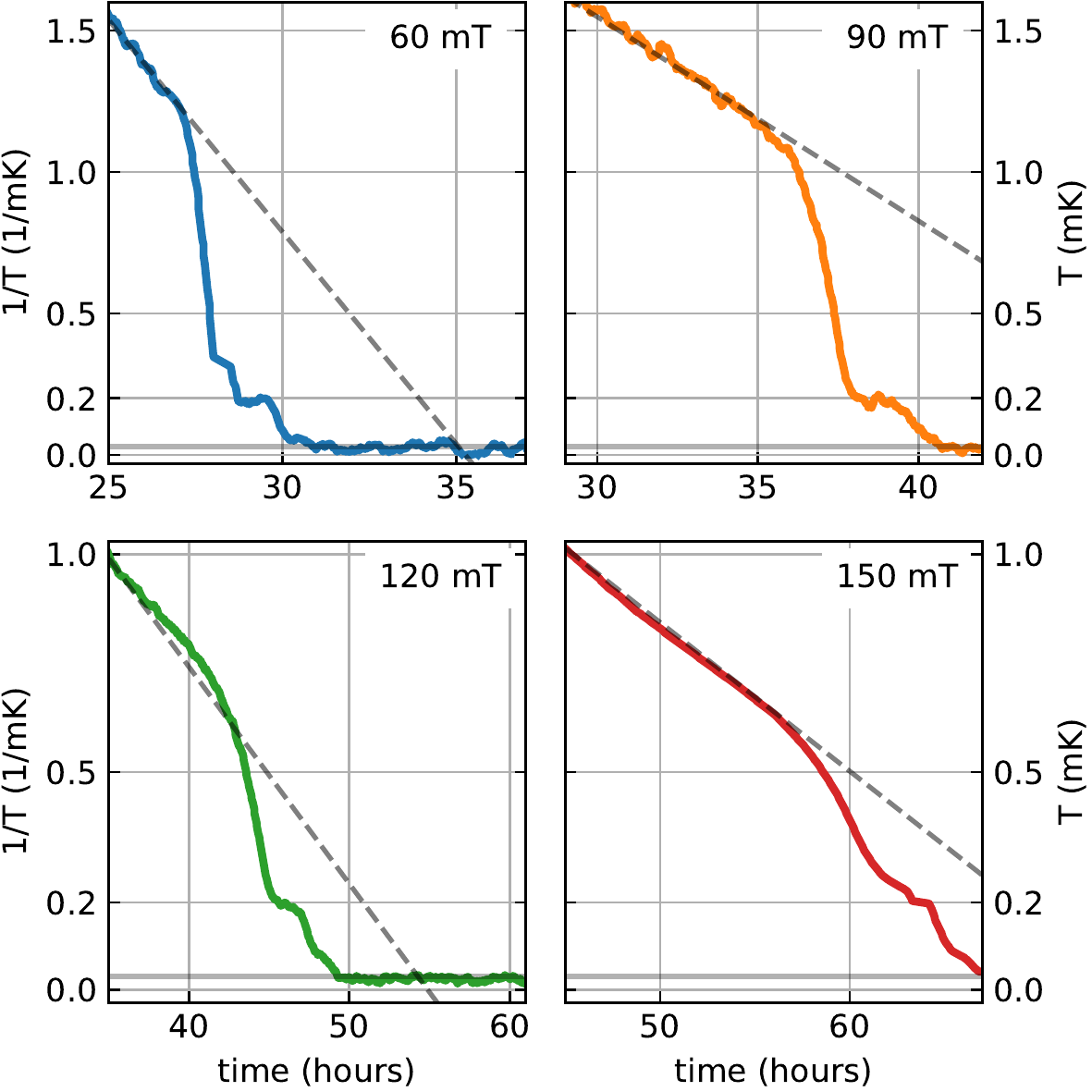}
	\caption{{\bf Rapid warmups at different fields} Zoomed in on the quick warmup steps (solid curves) as well the tails of the linear fits (dashed) in Fig.~\ref{fig:heatleaks}(a). A step at 5~mK is visible for warmups in all final fields.}
	\label{fig:heatleaks_zoomed}
\end{figure}

After replacing the loose wires inside the magnetic field with thicker and more rigid wires, the heat leak into the CBT is reduced by roughly an order of magnitude. As a consequence, the hold time drastically increases and is now limited by the macroscopic plates. It is worth investigating in the future whether turning off the pulse tube during the warmup still has an effect on this low heat leak after it has been reduced.

\begin{figure}[ht]
	\centering
	\includegraphics[width=\linewidth]{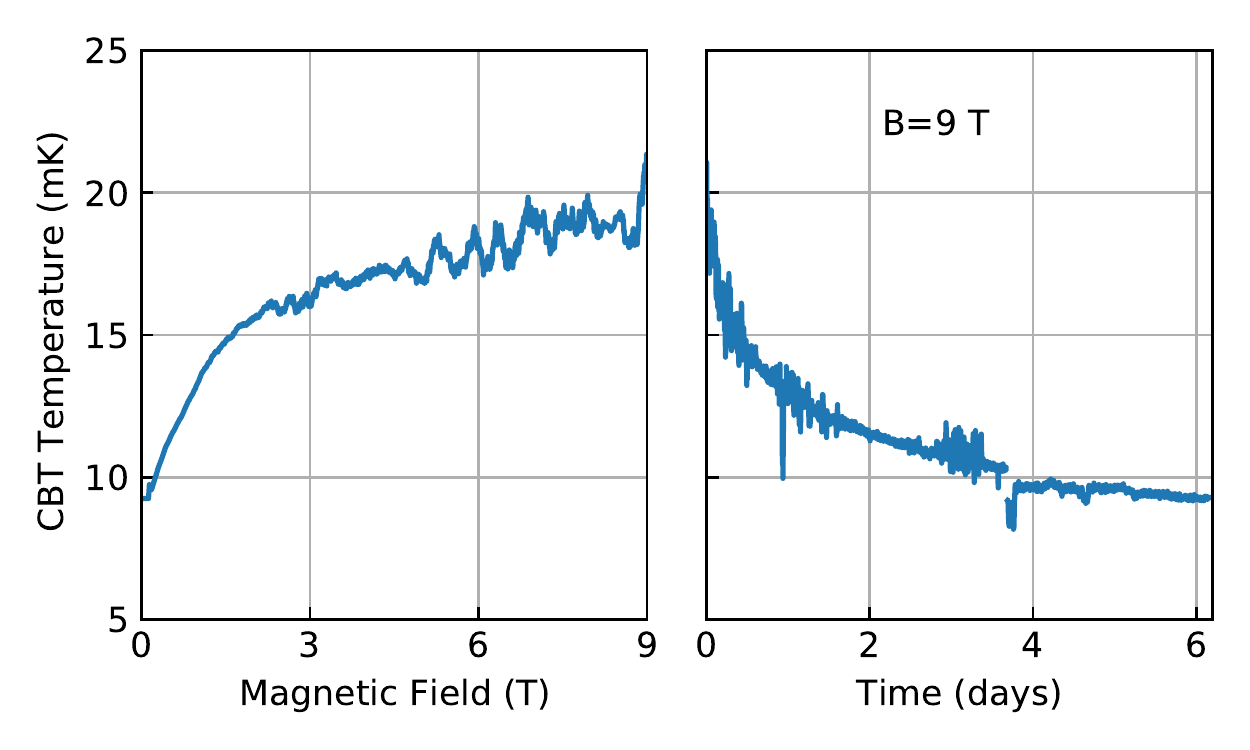}
	\caption{{\bf Precooling} Left: the heat of magnetization raises the temperature as the magnetic field is ramped up. Right: over the course of 6 days, the mixing chamber absorbs this heat.}
	\label{fig:precool}
\end{figure}

The new sample box also improved the precooling efficiency significantly. In Fig.~\ref{fig:precool}, we show the longest precooling with the new sample box. Only after 6 days, heat of magnetization is completely absorbed by the mixing chamber and the temperature of the CBT at 9~T is as low as temperature at low field before ramping up the magnet.

In Fig.~\ref{fig:demag} in the main article, the temperature rises rapidly from around 300~µK to a few mK. Comparing the rapid part of the warmup for different final magnetic fields (Fig.~\ref{fig:heatleaks}(a) and the zoomed-in version in Fig.~\ref{fig:heatleaks_zoomed}) reveals a two-step warmup, first to about 5~mK and later to $\approx$30~mK.


During the warmup, if the sample box, bolted to a macroscopic copper plate, warmed up first, the heat would leak through the top gate of the CBT via phonon-coupling and warm up the CBT very quickly and to high temperatures, above 10~mK. This is not what we observe, and thus the sample box does not warm up first. Presumably, the bond pads warm up first, introducing rather ineffective Wiedemann-Franz heating via the resistive array of tunnel junctions in the CBT. Any remaining cooling power of the islands and sample box keep the CBT at $\approx$5~mK for about two hours, forming a short plateau. Finally, the sample box (and macroscopic Cu plate connected to it) also warms up and the CBT reads a very high temperature, $\approx$30~mK.

The bond pads are precooled primarily via the sample box though varnish and the 5~m$\Omega$ microwave filters, thus saturating at slightly higher precooling temperature than the sample box itself which is connected electrically to the macroscopic copper plates. As a consequence, the copper box reaches a lower final temperature after demagnetization than the bond pads and hence has a longer hold time, consistent with the bond pads warming up before the copper plates and the  sample box.

\end{document}